\documentstyle[preprint,tighten,aps]{revtex}

\begin{document}
\draft
\title{A Network of Oscillators for Retrieving Phase Information}
\author{Toshio Aoyagi}
\address{Department of Applied Mathematics and Physics, Kyoto University,
Kyoto, Japan}
\date{\today}
\maketitle
\begin{abstract}
We propose a network of oscillators to retrieve given patterns in which
the oscillators keep a fixed phase relationship with one another.
In this description, the phase and the amplitude of the oscillators can be
regarded as the timing and the strength of the neuronal spikes,
respectively.
Using the amplitudes for encoding, we enable the network to realize
not only oscillatory states but also non-firing states.
In addition, it is shown that under suitable conditions the system has a
Lyapunov function ensuring a stable retrieval process.
Finally, the associative memory capability of the network is demonstrated
numerically.
\end{abstract}

%

Although the past decade has seen considerable advances in studies of
neural networks,
recent research seems to reveal the limitation of the current network
models, composed of McCulloch-Pitts units or modifications of these.
Provided that we will study the steady states of the network,
these units are valid for modeling of actual neurons as a first approximation.
In fact, many fruitful results have been reported by using these
units \cite{Hopfield}.
However, when it comes to treating the dynamical behavior of the network,
these models may fail to capture the essence of the dynamics.
This is because these models ignore much of the detailed behavior of real
neurons, such as the timing of neuronal firing and internal dynamics
\cite{Abbot,Aoyagi,Kuramoto}.
In some cases, such behavior seems to play a significant role
in neuronal systems.

For example, several recent experiments suggest that the temporal coherence of
neuronal activity may contribute to segmentation of visual scenes\cite{Gray}.
In this mechanism, synchronization of pulses of neuronal responses plays a
crucial role in binding the visual features belonging to the same object.
In other words, the phase relationships among the pulses become essential for
information processing.
Central pattern generators provide another good example.
It is widely believed that a well-identified group of neurons controls
rhythmic behavior of animals, i.e. locomotion, swimming, breathing and so on.
Such a neuronal group is often called a central pattern generator.
Particularly in case of an animal locomotion, it is known that there are
generally several distinct patterns of leg movements\cite{Stewart}.
These patterns are classified by the relative phase relationships
among the legs.
Therefore, it is obvious that the corresponding neuronal system can generate
several firing patterns for which there are different phase relationships among
the neuronal pulses.


To construct a theoretical model of the above systems,
we need to describe the temporal features of neuronal activities, such as
synchronization and phase locking.
However, a McCulloch-Pitts description is based on the assumption that
information is encoded only by the averaged activities of the neurons.
Therefore, such a description is too crude to represent temporal features
of the firing states naturally.
For example, if information is encoded by the relative phases of the spikes,
this situation cannot be described naturally by the McCulloch-Pitts units.
Hence, we need to construct another model which provides a suitable
framework to grasp such temporal aspect of real neural networks.
At the same time, it is required that such a model be simple enough to be
mathematically tractable.
For this purpose, a model in which neuronal activities take the form of
oscillators is an attractive candidate\cite{Strogatz,Fukai,Sompolinsky}.
Using this idea, we can treat the timing of neuronal firings
as the phase of the oscillators naturally.
This is because the firing state can be treated as a limit cycle
of a dynamical system where the time of emission of a pulse is indicated
by the position on the limit cycle, namely, the phase $\phi$.
Moreover, it is desirable that the non-firing state can also be expressed
by the model.
It is natural for the amplitude $r$ of the limit cycle to be thought of
as the strength of the pulse.
Pursuing this idea,
the non-firing state can be represented by the zero amplitude of the
oscillator.
For the neuron model, therefore, we will adopt an oscillatory unit whose state
is determined by both the phase and the amplitude.
For convenience of expression, we denote the state of the i-th neuron
by the complex variable $W_i$ with amplitude $r_i$ and phase $\phi_i$.
In this description, a non-firing state of the neuron corresponds to $W_i=0$,
while a firing state corresponds to $W_i=\exp(i\Omega_i t)$
($\Omega_i$ is the frequency of the i-th firing neuron.)


Now we will construct a neural network model to retrieve phase information.
Let us consider a network of $N$ neurons whose dynamics are governed by
the following system of coupled oscillators
\begin{equation}
{dW_i\over dt}=v(W_i, \tilde W_i)+k(\sum_{j=1}^N C_{ij}W_j -W_i),
\label{model1}
\end{equation}
where the real variable $k$ represents the total coupling strength, the
complex variable $C_{ij}$ represents the effect of the interaction between
the $j$-th and $i$-th neurons, and $\tilde W_i$ denotes the complex conjugate
of $W_i$.
The function $v(W_i, \tilde W_i)$ should be chosen so that the system can
exhibit limit cycle behavior in the absence of the coupling term.
In addition, we assume that the system (\ref{model1}) is invariant under
uniform phase translation
\begin{equation}
W_i \rightarrow W_i \exp(i\phi_0),
\label{trans}
\end{equation}
where $\phi_0$ is an arbitrary real constant.
This requirement arises from the fact that information is encoded not by
the absolute time but by the relative time of neuronal spikes.
Correspondingly, the relative phase relationships among the oscillators
are relevant to encoding the information (not the absolute phase values).
In this paper, for $v(W_i, \tilde W_i)$ we will examine two types of
functions satisfying the above two requirements.

We here define notation for later discussion.
Let $\xi_i^\mu$ $(\mu=1,\dots,p)$ be a set of patterns
to be memorized, where $p$ is the total number of the patterns.
We should remark that, owing to the invariance (\ref{trans}),
all patterns generated by the uniform phase translation
$\xi_i^\mu\exp(i\phi_0)$ represent the same pattern as $\xi_i^\mu$.
We also define {\it $p$ overlaps $M_\mu$} as the projections of the current
state on the $p$ embedded patterns:
\begin{equation}
M_\mu={1\over N}\left|\sum_{j=1}^N \tilde\xi_i^\mu W_j \right|.
\end{equation}

At first, we consider the situation that all neurons are in the
firing state. Stored patterns can then be characterized by the parameters
$\theta_i^\mu$ ($\xi_i^\mu=\exp(i\theta_i^\mu)$).
Due to (\ref{trans}), the effective number of degrees of freedom of these
parameters is $N-1$ for one pattern.
As a simple choice, we consider the following dynamics of the first model
\begin{equation}
{dW_i\over dt}=(1+i\Omega_i)W_i-(1+ic)|W_i|^2 W_i
+k(\sum_{j=1}^N C_{ij}W_j -W_i),
\label{gmodel}
\end{equation}
\noindent
where $\Omega_i$ can be considered as natural frequencies of neuronal spikes
in the absence of external inputs, $k$ controls the total strength of the
coupling, and $c$ is a real parameter.
The system (\ref{gmodel}) without the coupling terms is often referred as
the Stuart-Landau equation.
In general, it is well known that the Stuart-Landau equation is derived
from a general ordinary differential equation near a Hopf
bifurcation point.
Therefore, this choice seems to be natural for oscillatory systems.
For simplicity, we assume that $c=0$ and $\Omega_i=\Omega$.
Under this assumption, we can eliminate $\Omega$ by the transformation
$W_i\rightarrow W_i\exp(i\Omega t)$.
After simple calculations, therefore, (\ref{gmodel}) reduces to
\begin{equation}
{dW_i\over dt}=W_i-|W_i|^2 W_i +k(\sum_{j=1}^N C_{ij}W_j -W_i).
\label{pmodel}
\end{equation}

Next we will show how given patterns are embedded in the network of
oscillators, that is, how to determine the synaptic efficacies.
At first, let us consider the condition that the equation (\ref{pmodel})
has the solution $W_i=\xi_i^\mu$.
Substituting $W_i=\xi_i^\mu$ into (\ref{pmodel}) and using
$|\xi_i^\mu|=|\exp(i\theta_i^\mu)|=1$,
we obtain $\sum_{j=1}^N C_{ij}\xi_j^\mu=\xi_i^\mu$.
This condition can be rewritten in matrix form as
\begin{equation}
{\bf CP}={\bf P},
\label{scnd}
\end{equation}
where the $N\times P$ matrix ${\bf P}$ is defined by $P_{ij}=\xi_i^j$.
Provided that the total number $p$ of the memorized patterns is smaller
than $N$, the synaptic efficacies satisfying the condition (\ref{scnd}) are
then given by
\begin{equation}
{\bf C}={\bf PP}^{\dag}+{\bf B}({\bf I-PP}^{\dag}),
\label{gc}
\end{equation}
where ${\bf P}^{\dag}$ denotes the pseudo-inverse of ${\bf P}$, ${\bf I}$
denotes the unit matrix, and ${\bf B}$ is an arbitrary
matrix\cite{Personnaz,Kohonen}.
If the patterns are linearly independent,
the pseudo-inverse matrix ${\bf P}^{\dag}$ is given by
${\bf P}^{\dag}=1/N\left(1/N\tilde {\bf P}^T
{\bf P}\right)^{-1}\tilde {\bf P}^T$,
where $\tilde {\bf P}$ and ${\bf P}^T$ denote the complex conjugate and
the transpose of ${\bf P}$, respectively.
Moreover, assuming ${\bf B}={\bf O}$ for simplicity, we obtain
\begin{mathletters}
\label{synapse}
\begin{equation}
C_{ij}={1\over N}\sum_{k=1}^p \sum_{l=1}^p (A^{-1})_{kl}
\xi_i^k\tilde\xi_j^l,
\end{equation}
\begin{equation}
A_{ij}={1\over N}\sum_{k=1}^N \xi_k^i\tilde\xi_k^j ,
\end{equation}
\end{mathletters}
\noindent
with $\xi_i^\mu=\exp(i\theta_i^\mu)$.
It is easily found that the synaptic matrix (\ref{synapse}) is Hermitian,
that is, $C_{ij}=\tilde C_{ji}$.
It is worth noting that if the patterns are orthonormal the synaptic
prescription (\ref{synapse}) recovers the generalized Hebbian rule
$C_{ij}=1/N \sum_{k=1}^p \xi_i^\mu \tilde\xi_j^\mu$.

There are two trivial cases in this model.
For $p=1$, the synaptic efficacies are given by
$C_{ij}=1/N\exp(i\theta^1_i-i\theta^1_j)$.
In this case, applying the transformation $W_i\exp(-i\theta_i)\rightarrow W_i$,
the dynamics of (\ref{pmodel}) reduce to
\begin{equation}
{dW_i\over dt}=W_i-|W_i|^2 W_i +k({1\over N}\sum_{j=1}^N W_j -W_i).
\end{equation}
This equation has recently been studied by several
authors\cite{Nakagawa,Hakim}.
It is thought that this situation corresponds to the Mattis state in spin glass
models.
On the other hand, for $p=N$ we obtain $C_{ij}=\delta_{ij}$.
This implies that the couplings between the oscillators vanish and
the dynamics of the network reduces to that of $N$ independent oscillators.

We will show that the system (\ref{gmodel}) reduces to a simpler set of
phase oscillators under the conditions that the coupling is weak and $c=0$.
In the limit of weak coupling $k \sim O(\epsilon)$,
the deviation of the amplitude $r_i$ from $r=1.0$ may be neglected
at first order.
Inserting $W_i=r_i \exp (i\phi_i)$ into (\ref{pmodel}),
this approximation immediately yields
\begin{equation}
\dot \phi_i=\Omega_i+k\sum_j |C_{ij}|\sin(\phi_j-\phi_i+\alpha_{ij}),
\label{phmodel}
\end{equation}
where the $\alpha_{ij}$ are defined by $C_{ij}=|C_{ij}|\exp(i\alpha_{ij})$.
Since the synaptic matrix is Hermitian, we get
\begin{equation}
|C_{ij}|=|C_{ij}| \hskip 0.5cm \hbox{and} \hskip 0.5cm
\alpha_{ij}=-\alpha_{ji}.
\label{hermit}
\end{equation}
Note that $\alpha_{ij}=-\alpha_{ji}$ leads $\alpha_{ii}=0$.
As a result, self-coupling terms do not contribute to the dynamics of the
network.


In the previous model (\ref{gmodel}), all neurons are assumed to exhibit
periodic firing states. However, such an assumption is not realistic because
in a real system, for any given patterns, some neurons will be in a resting
state.
In the second model, we consider another type of dynamics where the amplitude
plays a crucial role in representing non-firing state.
Let us first consider the dynamics of a single neuron.
To realize a stable non-firing state in retrieval patterns,
it is desirable that the single neuron is able to exhibit both a periodic
firing state and a non-firing state.
To satisfy this requirement, we choose $v(W_i,\tilde W_i)$ so that the
dynamics of a single neuron can have both a limit cycle $W_i=\exp(i\Omega_i t)$
and a stationary fixed point $W_i=0$.
As a simple choice, we employ
$v(W_i,\tilde W_i)=(-1+i\Omega_i)W_i+4|W_i|^2 W_i-3|W_i|^4 W_i$.
As is the case of the first model,
assuming that the natural frequencies $\Omega_i$ are identical to $\Omega$,
we can eliminate $\Omega$.
Consequently, the dynamics of the network are described by the following
\begin{equation}
\dot W_i=-W_i+4|W_i|^2 W_i-3|W_i|^4 W_i+k\left( \sum_{j=1}^N C_{ij}
W_j -W_i \right)
\label{model2}.
\end{equation}

We next address the question of how to make the synaptic connections for
patterns which include the possibility of non-firing states.
The patterns are then defined by
\begin{equation}
\xi_i^\mu=\left\{
\begin{array}{ll}
\exp(i\theta_i^\mu)\hskip 1cm&\hbox{for firing state}\\
0&\hbox{for non-firing state}
\end{array}
\right.
\end{equation}
Putting the above patterns into (\ref{model2}) and using $|\xi_i^\mu|=1$ or
$0$, we obtain the same condition for the synaptic connections,
${\bf CP}={\bf P}$, as in the first model.
Therefore, if $p<N$ and the patterns are linearly independent,
we can apply the same procedure as in the first model.
As a result, the synaptic matrix is given by the same prescription
(\ref{synapse}).
Note that the condition $C_{ij}=\tilde C_{ji}^T$ is kept even if the memorized
patterns include some non-firing states.


Above, it was shown that we can set the connections so as to
make the pattern to be memorized the solution of the dynamical equation.
To recall the embedded patterns from noisy ones, however, it is required that
such solutions are the attractors of the dynamics.
The existence of such asymptotic behavior is guaranteed by the existence of a
Lyapunov function.
In this section, we will show that in our models a Lyapunov function
exists if the synaptic matrix is Hermitian and $\Omega_i=\Omega$.
The models which we have discussed so far can be written in the form
\begin{equation}
{dW_i\over dt}= v(W_i,\tilde W_i)+k\left( \sum_{j=1}^N C_{ij}W_j -W_i \right),
\label{g2model}
\end{equation}
with
\begin{equation}
v(W_i,\tilde W_i)=
\left\{\begin{array}{l}
W_i-|W_i|^2 W_i \\
-W_i+4|W_i|^2 W_i-3|W_i|^4 W_i .
\end{array}
\right.
\label{vfunc}
\end{equation}
Let us introduce a function $V(W_i,\tilde W_i)$ by the following
definition
\begin{equation}
v(W_i,\tilde W_i)=-{\partial V(W_i,\tilde W_i) \over \partial \tilde W_i}
\quad \hbox{and} \quad
\tilde v(W_i,\tilde W_i)=-{\partial V(W_i,\tilde W_i) \over \partial W_i},
\end{equation}
where we have regarded $W_i$ and $\tilde W_i$ as independent variables.
For example, the corresponding functions $V(W_i,\tilde W_i)$ for
(\ref{vfunc}) are given by
\begin{equation}
V(W_i,\tilde W_i)=
\left\{\begin{array}{l}
-|W_i|^2+{1\over 2}|W_i|^4\\
|W_i|^2-2|W_i|^4+|W_i|^6 .
\end{array}
\right.
\end{equation}
Provided that such a function $V(W_i,\tilde W_i)$ exists,
the Lyapunov function is given by
\begin{equation}
L(W_i,\tilde W_i )= \sum_{i=1}^N V(W_i,\tilde W_i)
-{k\over 2}\sum_{i=1}^N \sum_{j=1}^N \left(C_{ij}\tilde W_i W_j +
\tilde C_{ij} W_i \tilde W_j\right)
+k\sum_{i=1}^N|W_i|^2 .
\end{equation}
\noindent
It is easily proved that $L$ only decreases under the dynamics of
(\ref{g2model}).
Using the function $V(W_i,\tilde W_i)$,
we can rewrite the model (\ref{g2model}) in the from
\begin{equation}
{dW_i\over dt}= -\left(\partial L\over \partial \tilde W_i \right)
\quad \hbox{and} \quad
{d\tilde W_i\over dt}= -\left(\partial L\over \partial W_i \right) .
\end{equation}
{}From these relations, we immediately get
\begin{equation}
\begin{array}{rl}
\displaystyle{dL\over dt}&=
\displaystyle\sum_i^N \left( {\partial L\over\partial W_i}
{dW_i\over dt}+{\partial L\over\partial \tilde W_i}{d\tilde W_i\over dt}
\right)\\
&=-\displaystyle\sum_i^N \left|
v(W_i,\tilde W_i)+k\left( \sum_{j=1}^N C_{ij}W_j -W_i \right)
\right|^2 \\
&\leq 0 .
\end{array}
\end{equation}
Therefore, $L$ can only decrease as a function of time.
When the network converges to the memorized state $W_i=\xi_i^\mu$,
$L$ does not vary with time.
We should remark that $L$ is invariant under the uniform phase translation
(\ref{trans}).



To confirm the ability of the network, we carried out some numerical
simulations. Among these simulations, we present one typical result.
In this simulation, we used a network of 50 oscillators whose dynamics are
govern by the equations (\ref{model2}).
The reason for using the second model is that we want to demonstrate the
ability of the network to retrieve phase patterns which include non-firing
states.
In all, eight patterns were stored by means of the synaptic prescription
(\ref{synapse}). The first pattern was $\xi^1=(1,1,1,1,0,0,1,1,1,1,$
$e^{i{2\pi\over 5}},e^{i{2\pi\over 5}},e^{i{2\pi\over 5}},e^{i{2\pi\over 5}},0,
0,
e^{i{2\pi\over 5}}, e^{i{2\pi\over 5}}, e^{i{2\pi\over 5}},e^{i{2\pi\over
5}},\cdots,
e^{i{8\pi\over 5}},$
$e^{i{8\pi\over 5}},e^{i{8\pi\over 5}}, e^{i{8\pi\over 5}},
0,0,e^{i{\pi\over 5}},e^{i{8\pi\over 5}},e^{i{8\pi\over 5}}, e^{i{8\pi\over
5}})$.
The remaining seven patterns $\xi_i^\mu$ $(p=2 \sim 8)$ were generated at
random. That is, taking the form $\xi_i^\mu=A_i^\mu \exp(i\theta_i^\mu)$,
$\theta_i^\mu$ were chosen at random from a uniform distribution
between $0$ and $2\pi$, and $A_i^\mu$ were independent random variables
obeying the distribution
\begin{equation}
A_i^\mu=\left\{
\begin{array}{ll}
1\hskip 1cm&\hbox{with probability $1\over 5$}\\
0&\hbox{with probability $4\over 5$}
\end{array}
\right.
\end{equation}
Figure 1 shows one typical temporal pattern of the retrieval process in which
the parameter values are $k=1.0, \Omega=2\pi$.
Although $\Omega$ make no contribution to the dynamics of the network,
to express the phase relationships visually,
the pattern in Figure 1 is illustrated in the rotating frame.
In this context, the black horizontal bars represent active phases defined
by Re$W_i>0.5$.
The network was initially given the noisy pattern $\xi_i^1$.
Obviously, both the amplitudes and the phases were corrected dynamically.
In the same simulation, time development of overlaps is shown in Figure 2.
It is clearly found that the network succeeds in retrieving one of
eight memorized patterns.
The other simulation results support the generality of the retrieval ability
in our model.
Therefore, we confirmed that the network of oscillators proposed in this
paper works well for associative memory of phase patterns.


We would like to make some comments here before concluding.
First, the synaptic efficacies, $C_{ij}$, was regarded as complex numbers.
This may be naturally explained by the fact that the neurons are coupled
via more than one component\cite{Ermentrout}.
Secondly, even if the natural frequencies $\Omega_i$ have a nontrivial
distribution, it is expected that the retrieval of the patterns will be
achieved by the network.
By analogy with a network of phase oscillators, it may then be required that
$k$ assumes a value larger than a certain critical value $k_c$.
Finally, it is well known that the storage capacity of an oscillator network
constructed by using the Hebbian rule is given by
$\alpha_c=P/N=0.0377$\cite{Cook,Gerl,Okuda}.
Therefore, using the generalized Hebbian rule, we expect that our models work
well when $\alpha <\alpha_c$.


In conclusion, we proposed a network of oscillators for the retrieval of phase
information. We showed that this network has the ability to retrieve given
patterns in which the oscillators keep a fixed phase relationship with one
another.
We would like to emphasize that, in the case of the second model, the network
can retrieve patterns which include non-firing states.
The amplitudes of the oscillators then play a crucial role in representing
non-firing states.
Furthermore, it is shown that under suitable conditions the system has a
Lyapunov function ensuring a stable retrieval process.
Using numerical simulations, we confirmed the good performance of
our models.
Consequently, we believe that the proposed models serve as a convenient
starting point for the study of oscillatory neuronal systems.

\acknowledgments
I would like to thank Y. Kuramoto, K. Okuda, T. Chawanya and
I. Nishikawa for fruitful discussions and suggestions.
I also thank G. Paquette for critically reading the manuscript.
This work is supported by the Japanese Grant-in-Aid for Science
Research Fund from the Ministry of Education, Science and Culture
(Nos. 06260102 and 06740321).


%


\begin{figure}
\caption{
Typical temporal pattern of retrieval process in the second model.
Black  horizontal bars show active phases which are defined by
$\mbox{Re} W_i>0.5$. The network succeeds in retrieving one of
8 memorized patterns.
Note that the memorized pattern includes non-firing states.
}
\label{fig1}
\end{figure}


\begin{figure}
\caption{
Time development of eight overlaps in the same simulation as Figure 1.
The solid line shows the overlap $M_1$ concerning the retrieval pattern.
}
\label{fig2}
\end{figure}


\begin{references}

\bibitem{Hopfield}
J.J. Hopfield, {\it Proc. Nat. Acad. Sci. USA} {\bf 79}, 2554 (1982).
\bibitem{Abbot}
L.F. Abbot, {\it J.Phys.} {\bf A23}, 3835 (1990).
\bibitem{Aoyagi}
T. Aoyagi, {\it Europhys. Lett.} {\bf 20}, 565 (1992) .
\bibitem{Kuramoto}
Y. Kuramoto, T. Aoyagi, I. Nishikawa, T. Chawanya and K. Okuda,
{\it Prog. Theor. Phys.} {\bf 87}, 1119 (1992).
\bibitem{Gray}
C.M. Gray, P. K\"onig, A.K. Engel and W. Singer, {\it Nature} {\bf 338},
334 (1989).
\bibitem{Stewart}
J.J. Collins and I.N. Stewart, {\it to appear in the Journal of Nonlinear
Science}.
\bibitem{Strogatz}
P.C.Matthews, R.E.Mirollo and S.H.Strogatz, {\it Physica} {\bf D52}, 293
(1991).
\bibitem{Fukai}
T. Fukai, and M. Shiino, {\it Europhys. Lett.} {\bf 26}, 647 (1994).
\bibitem{Sompolinsky}
H. Sompolinsky, H. Golomb and D. kleinfeld, {\it Phys.Rev.} {\bf A43},
6990 (1991).
\bibitem{Personnaz}
L.G. Personnaz, I. Guyon and G. Dreyfus,{\it Phys. Rev.} {\bf A34}, 4217
(1986).
, {\it J. Physique Lett.} {\bf 46}, L359 (1985).
\bibitem{Kohonen}
T. Kohonen, {\it Self-Organization and Associative Memory}
(Springer-Verlag, 1984).
\bibitem{Nakagawa}
N. Nakagawa and Y. Kuramoto, {\it Prog. Theor. Phys.} {\bf 89}, 313 (1993).
\bibitem{Hakim}
V. Hakim, and W. J. Rappel, {\it Phys. Rev.} {\bf A46}, 46 (1992).
\bibitem{Ermentrout}
B. Ermentrout, and N. Kopell, {\it Neural comp.} {\bf 6}, 225 (1994).
\bibitem{Cook}
J. Cook, {\it J. Phys.} {\bf A22}, 2057 (1989).
\bibitem{Gerl}
F. Gerl, K. Bauer, and U. Krey, {\it Z. Phys.} {\bf B88}, 339 (1992).
\bibitem{Okuda}
K. Okuda, {\it unpublished}.
\end{references}
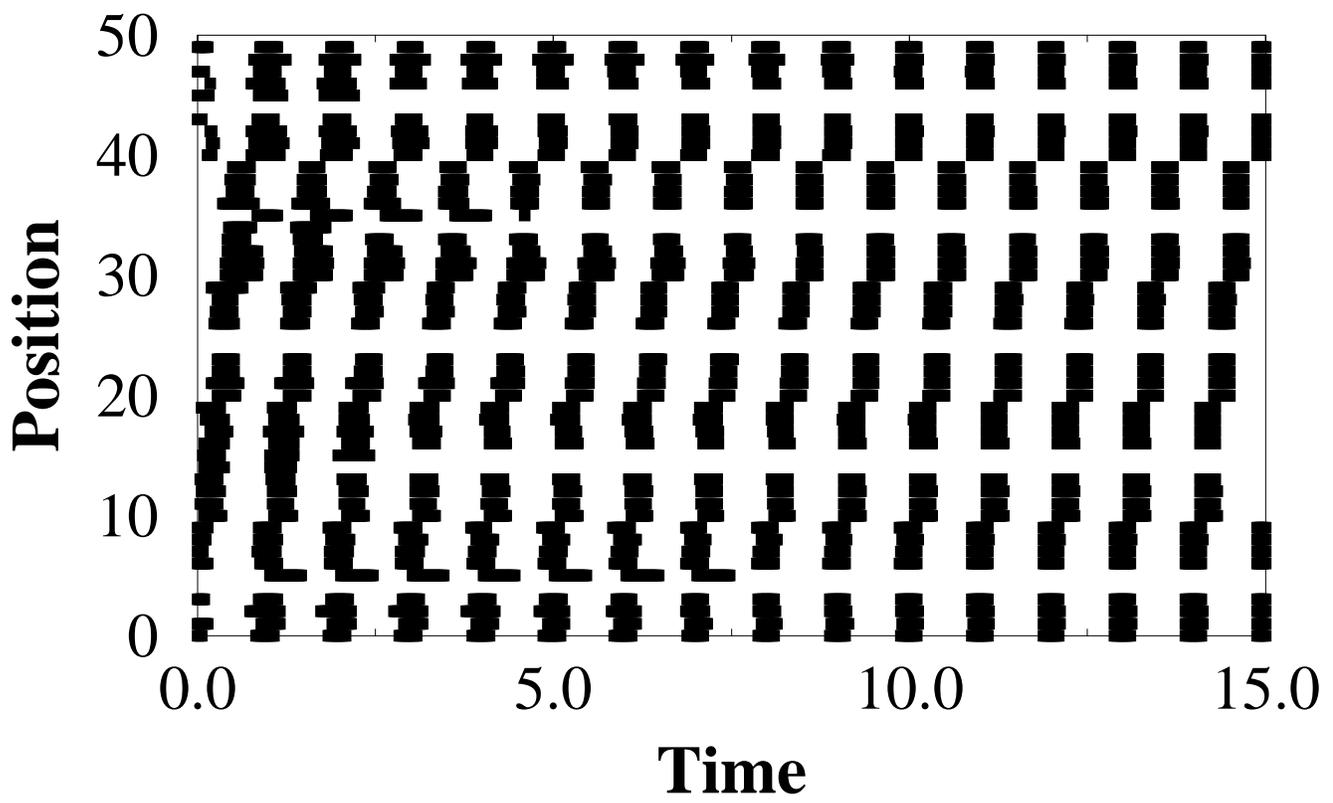
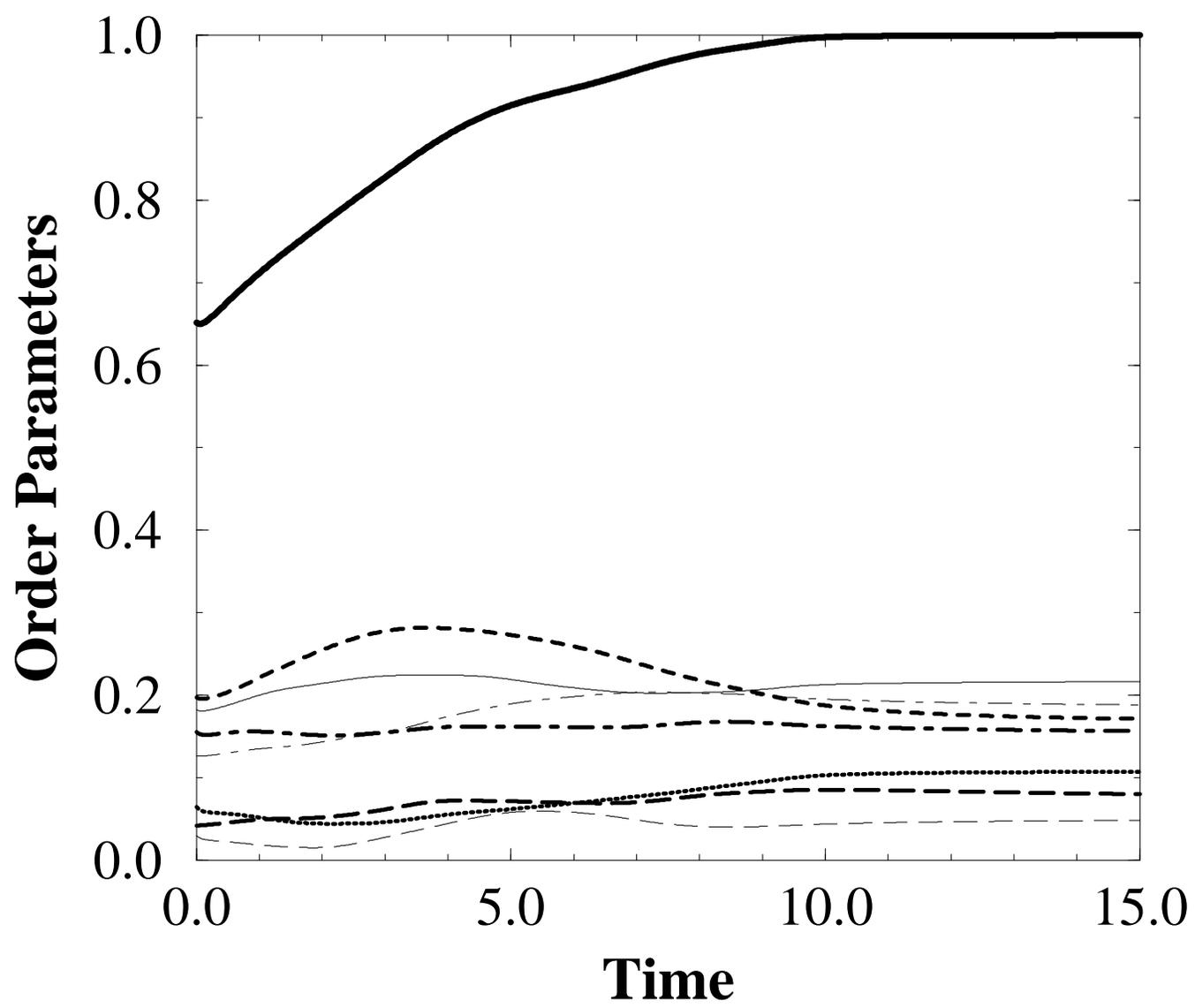
\end{document}